\documentclass[twocolumn,aps,prl,nobibnotes,8pt]{revtex4-1}
\usepackage{amssymb, amsmath, amsthm}
\usepackage{graphicx,import}
\usepackage{units}

\newcommand{\ket}[1]{\ensuremath{\vert{#1}\rangle}}

\usepackage{upgreek}
\usepackage{color}
\usepackage[usenames,dvipsnames]{xcolor}
\usepackage{url}
\usepackage{lineno}
\begin{document}
\title{Real-Time Near-Field Terahertz Imaging with Atomic Optical Fluorescence}
\author{C. G. Wade}\email{c.g.wade@durham.ac.uk}
\author{N. \v{S}ibali\'c}
\author{N. R. de Melo}
\author{J. M. Kondo}
\author{C. S. Adams}
\author{K. J. Weatherill}
\affiliation{Joint Quantum Centre (JQC) Durham-Newcastle, Department of Physics, Durham University, DH1 3LE United Kingdom}
\maketitle 


\noindent
{\bf
Terahertz (THz) near-field imaging is a flourishing discipline~\cite{adam11,chan07}, with applications  from fundamental studies of beam propagation~\cite{bitz08} to the characterisation of metamaterials~\cite{bitz09,acun08} and waveguides~\cite{mitr09,niel09}.
Beating the diffraction limit typically involves rastering structures or detectors with length scale shorter than the radiation wavelength; in the THz domain this has been achieved using a number of techniques including scattering tips~\cite{dean16,hube08} and apertures~\cite{bara11}.
Alternatively, mapping THz fields onto an optical wavelength and imaging the visible light removes the requirement for scanning a local probe, speeding up image collection times~\cite{wu96,doi10}.
Here we report THz to optical conversion using a gas of highly excited `Rydberg' atoms.
By collecting THz-induced optical fluorescence we demonstrate a real-time image of a THz standing wave and we use well-known atomic properties to calibrate the THz field strength.
}

\begin{figure*}[t]
\includegraphics[width=\textwidth]{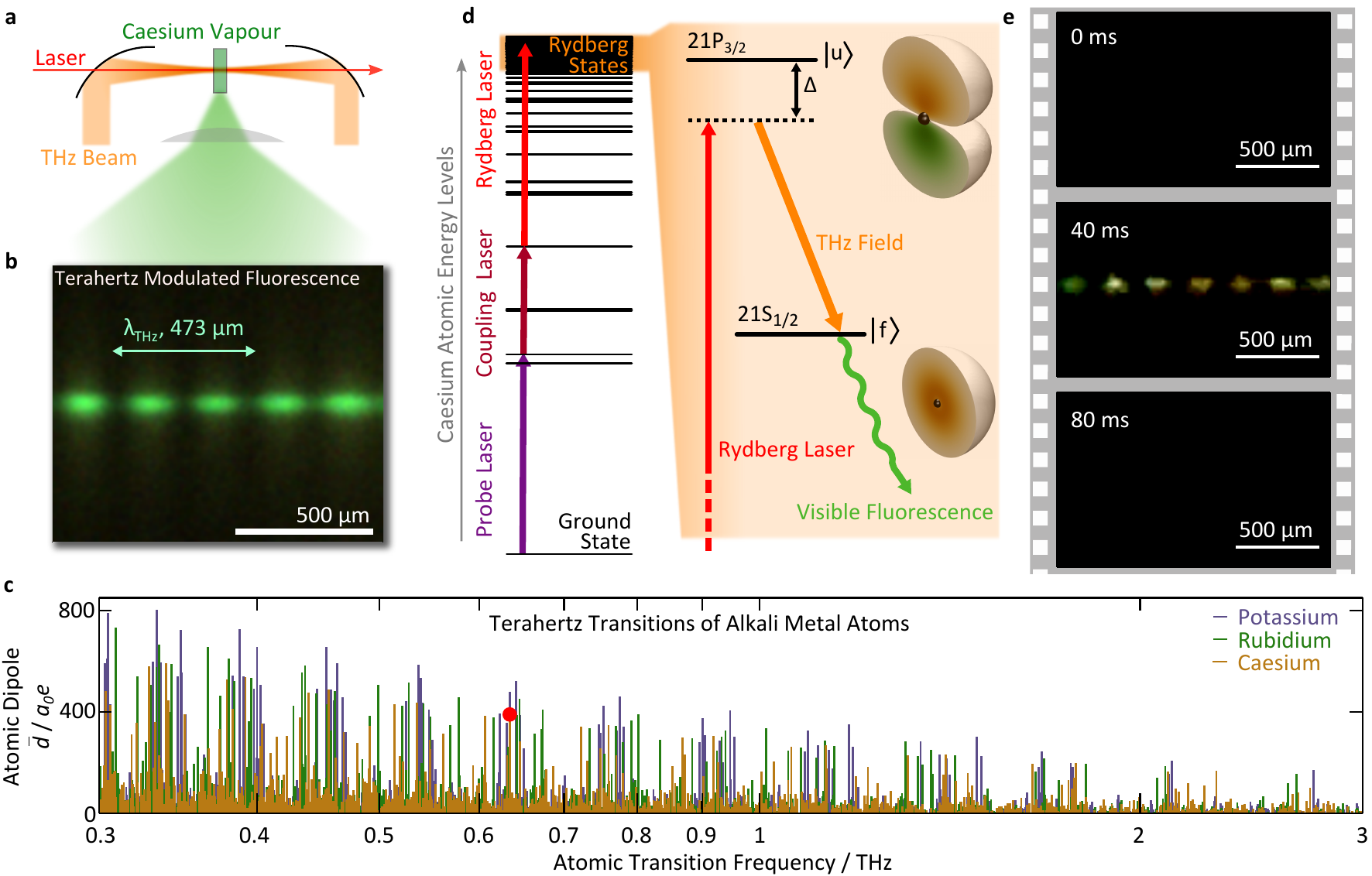} 
\label{fig:1}
\caption[]{THz Imaging with an Atomic Vapour: 
(a)~Experimental layout: The THz beam is co-linear with laser beams passing through a caesium vapour cell. 
(b)~Spatially Resolved THz Intensity: THz-induced atomic fluorescence (proportional to the THz intensity) is imaged by a consumer digital camera with a 0.5\,s exposure. The dark vertical stripes indicate the nodes of the THz standing wave, spaced at $\lambda$/2 intervals (237\,$\mu$m).
(c)~Rydberg transition frequencies: Resonant atomic transitions in potassium, rubidium and caesium span the THz gap, offering a wide selection of detection frequencies. We plot calculated values for the reduced dipole matrix elements. 
The red dot highlights the caesium transition used for this work.
(d)~Caesium atomic energy levels and laser excitation scheme: Three infra-red lasers combine with the THz field to excite atoms to the final Rydberg state, $\ket{f}$. The images reported in this letter were recorded with $\Delta/2\pi$~=~-243\,MHz. 
The THz-induced visible fluorescence emitted by the Rydberg atoms indicates the spatial profile of the THz field.
(e)~Real time video: The THz source is gated in synchrony with the video frames, revealing an image of the THz standing wave captured in a single 40~ms frame.
}
\end{figure*}

Atoms make excellent electromagnetic field sensors because narrow line-width atomic transitions couple strongly to EM fields, giving atoms a sensitive, narrowband response.
In addition, each atom of the same isotope is identical and has well studied, permanent properties which facilitate easy calibration to SI units.
Atomic states that couple to multiple transitions offer an interface between different frequency regimes.
In this way atomic ground states have been used to map microwave fields onto an optical probe~\cite{hors15}. 
However atomic ground states are only sensitive to a limited selection of microwave frequencies.
In contrast, highly-excited Rydberg atoms couple to strong, electric dipole transitions across a wide range of microwave and THz frequencies, making them ideal candidates for field measurement and for frequency standards in the millimeter wave and THz range~\cite{raim81}.
Previous methods for THz imaging with Rydberg atoms used the THz radiation to ionise the atoms~\cite{drab99,gurt03}. 
More recently optical read-out of Rydberg states was demonstrated in a room-temperature alkali-metal vapour using electromagnetically induced transparency (EIT)~\cite{moha07}.
The `Rydberg EIT' technique has since been exploited to readout radio frequency fields~\cite{moha08}, to demonstrate precision microwave electrometry~\cite{sedl12,gord14,simo16} and for sub-wavelength imaging of microwave fields~\cite{fan14}.

In distinction to the EIT technique we make direct use of THz-induced optical fluorescence to demonstrate THz imaging.
An overview of our THz imaging setup is shown in Figure~1a.
Infrared laser beams forming a three-step ladder excitation scheme~\cite{carr13b} are co-axially aligned with a continuous wave (CW) THz beam and pass through a caesium vapour in a 2~mm long quartz cell.
In regions where both the THz field and the laser beams are present, atoms are excited to a Rydberg state and subsequently decay with fluorescence at visible wavelengths.
The fluorescence is imaged by a consumer digital camera, and a typical 0.5\,s exposure is shown in Figure~1b. 
In Figure~1c we show THz transition frequencies and calculated reduced dipole matrix elements, $\bar{d}$, for alkali metal Rydberg atoms (see Supplementary Information for calculation details).
In principle any of these frequencies would be suitable for our detection technique with the sensitivity set by $\bar{d}$. 
We use the 0.634~THz transition between the 21P$_{3/2}$ and 21S$_{1/2}$ states of caesium (highlighted with a red dot), which matches the frequency of our THz source. 

The principle of operation of our THz-to-optical conversion scheme is a stimulated Raman transition~\cite{foot05}.
In order to prevent the creation of Rydberg atoms by laser excitation alone, the final laser is detuned from the upper Rydberg state, $\ket{u}$, by a frequency detuning $\Delta$ (Figure~1d).
Instead, Rydberg atoms are only created by the Raman transition which involves both the laser and THz fields.
The transition occurs when the THz field is detuned from the transition between $\ket{u}$ and a final Rydberg state $\ket{f}$ by the amount $\Delta$, matching the laser detuning.
Thus atoms are excited straight to state $\ket{f}$ at locations where the THz field and laser beams overlap in space.
We note that the THz field is not absorbed by the atoms because it drives stimulated emission.
The features in the camera image (Figure~1b) can be understood by considering the geometry of the laser and THz fields: both the laser and THz beams propagate horizontally across the image, but part of the THz beam is reflected back on itself to produce a standing wave interference pattern with nodes and anti-nodes perpendicular to the laser beam.
The fluorescence has the form of a horizontal line matching the width of the laser beams, with periodic modulation according to the THz intensity of the standing wave.

The image acquisition time is set by the fluorescence intensity, the camera sensitivity and the desired signal to noise ratio.
Using a generic consumer digital camera we record 25 frames per second video. 
In Figure~1e we show three consecutive video frames during which the THz source was gated in synchrony with the camera shutter so that alternating frames saw the field present or absent. 
In the second frame the standing wave is visible, while the first and third frames are blank. 
This video-rate imaging without averaging over repeated experimental runs establishes the real-time character of the imaging.
The underlying limit of the imaging bandwidth is likely set by the fluorescence lifetime of the Rydberg atoms which is on the order of microseconds.

\begin{figure*}[t]
\includegraphics[width=\textwidth]{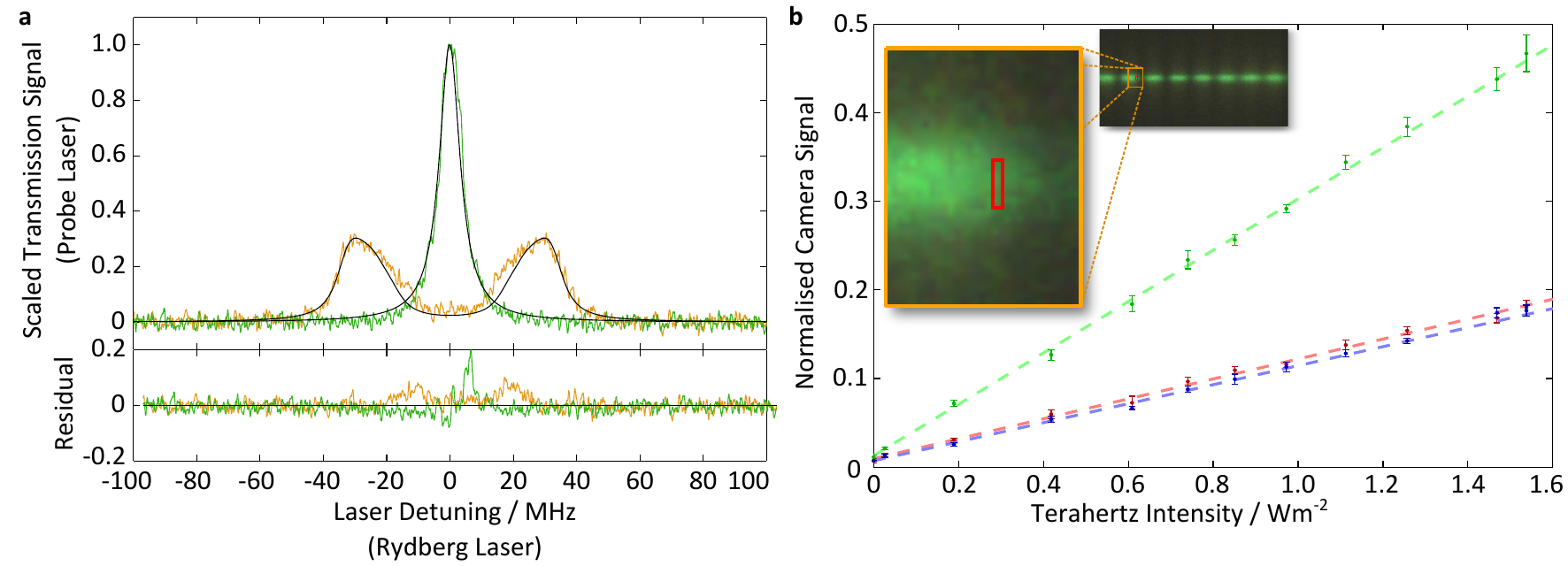}
\label{fig:2}
\caption[]{THz Field Calibration:
(a)~THz electric field measurement using electromagnetically induced transparency (EIT): 
The transmission of the probe laser reveals a narrow EIT feature (green) as the Rydberg laser is scanned. 
The feature is split into a doublet by the presence of a resonant THz field (orange). 
The peak splitting, $\Omega/2\pi$, is proportional to the THz electric field amplitude, and so the doublet peaks are broadened by the field variation in the standing wave.
The intensity variation from a matching image (Figure~3c) was used to fit the data (black line). 
(b)~Camera Calibration: We present the camera signal as a function of THz intensity.
The fluorescence signal and THz field intensity both relate to a small region of the image shown by the inset (red box).
The red, green and blue (rgb) data points and fit lines correspond to the rgb channels in the camera respectively. 
The error bars represent statistical variation in five repeated measurements.
}
\end{figure*}

For absolute calibration of the images we use the Rydberg EIT technique to make a separate measurement of the THz electric field amplitude along the laser beams~\cite{sedl12,carr13b}.
We measure the transmission of the probe laser (which is resonant with a transition from the ground state) as we scan the Rydberg laser (Figure~2a).
With the THz field on resonance ($\Delta=0$), Autler-Townes splitting divides a narrow Lorentzian feature into a doublet~\cite{autl55}, where the frequency interval between the peaks corresponds to the THz electric field, $E = \hbar \Omega / d$ where $\Omega/2\pi$ is the frequency interval and $d = \bar{d} / \sqrt{6}$ is the dipole matrix element between states $\ket{u}$ and $\ket{f}$ (see supplementary information). 
Spatial variation of the THz field along the laser beams leads to broadening of the Autler-Townes peaks. 
We can extract the relative variation of the THz field from a matching fluorescence image (Figure~3c), and use the variation to model to the lineshape (black line in Figure~2a - see supplementary information).
The only three free parameters in the model are the height of the peaks, the resonant laser frequency (zero detuning) and the scale of the THz electric field.

Using the Autler-Townes calibration technique we show that the pixel level in the image is proportional to the THz intensity (Figure~2b).
A small region of pixels is averaged to find the normalised camera signal (0~to~1~corresponds to the dynamic range of the camera).
We control the THz intensity with a well characterised attenuator at the THz source, and infer the local THz intensity  from the relative pixel level and the Autler-Townes spectrum.
When the THz field has zero-intensity, off-resonant laser excitation to state $\ket{u}$ causes weak background fluorescence, which we subtract from the images.
Once the sensitivity has been measured it is not necessary to repeat the calibration and background fluorescence measurement whilst recording real-time THz field images.


\begin{figure*}[t]
\includegraphics[width=\textwidth]{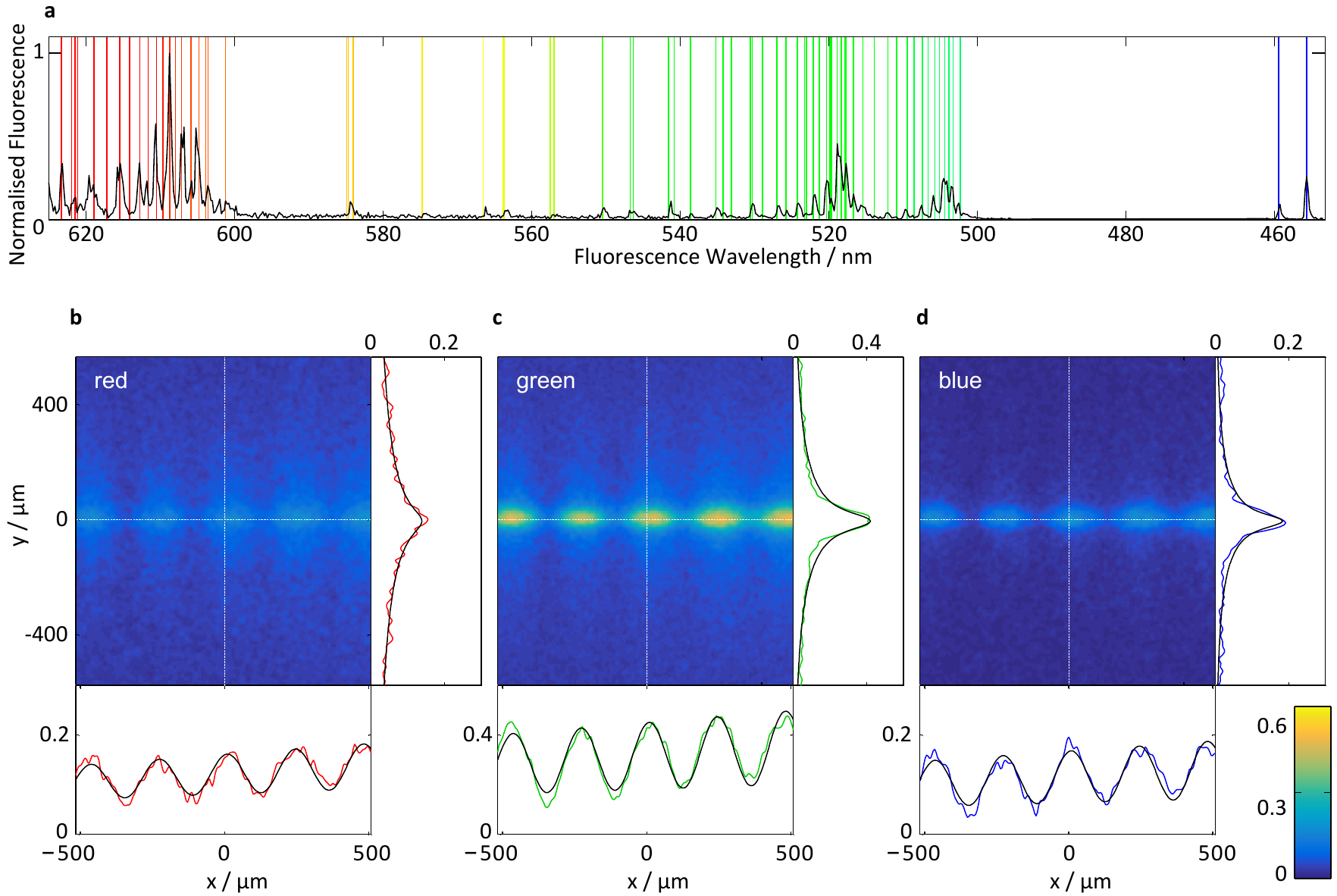}
\label{fig:3}
\caption[]{
(a)~Visible atomic fluorescence spectrum: The vertical lines indicate a selection of caesium transition wavelengths.
(b-d)~Colour channel analysis: Each colour channel of the image shown in Figure~1b records a particular set of optical decay pathways, giving different resolution according to the decay times.
Here we present the red, green and blue channels separately and show $x$ and $y$ cross sections of the camera channels at the positions indicated by white dotted lines.
}
\end{figure*}

Motion of the atoms in the time between excitation and decay results in slight blurring of the features in the image.
We are able to minimise this motion-induced blurring in the $x$-direction (along the laser beams) using Doppler selection, which suppresses the excitation of fast moving atoms.
The Doppler shift changes the laser frequency experienced by moving atoms according to the scalar product of the atom velocity, ${\bf v}$, and the vector sum, ${\bf k}$, of the laser wave-vectors.
Atoms moving in the $x$-direction parallel to ${\bf k}$ experience mis-matched laser frequencies which inhibit laser excitation, suppressing the contribution to the signal from fast moving atoms.
It is due to Doppler selection that we are able to see the interference fringes in the camera images. 
By crossing the excitation lasers in a light sheet configuration Doppler selection could be exploited in both image axes to make fast 2D measurements of THz fields.

The atomic fluorescence in the camera images includes emission lines that span the visible spectrum (Figure~3a).
Although the Raman transition only excites the 21S$_{1/2}$ state we see emission lines from many different Rydberg states, above and below in energy:
Red lines in the spectrum originate from Rydberg atoms in nP and nF states decaying to the 5D state;
the spectral lines present in the green channel encompass nS and nD atoms decaying to the 6P$_{3/2}$ and 6P$_{1/2}$ states, with the strongest contributions coming from nD states;
blue light originates from atoms decaying from the 7P to 6S ground state, the last step in a cascade process.
The complexity of the fluorescence spectrum originates from state transfer in the Rydberg manifold, though the dynamics of dense Rydberg vapours are not well understood~\cite{carr13a}. 
However, the fast Rydberg population transfer processes likely accelerate the optical decay and benefit the spatial resolution.

Each fluorescence line has a characteristic decay time and consequently each colour channel (red, green, blue) of the camera shows different resolution (Figures~3b-d).
We infer that the red channel corresponds to the slowest decay processes as it shows the smallest intensity fringe visibility, $V = 30$\% (extracted from a model for the THz intensity, $I = \big[1 - V{\rm cos}(2kx + \phi)\big](ax + b)$, with $k$ the wavevector of the THz wave, and parameters $a$, $b$ and $\phi$ fitted to the data).
The green and blue channels both show higher intensity fringe visibility (45\%) due to faster decay times.
The fringe contrast implies 24\% reflection of the THz electric field amplitude.
The spatial resolution could be optemised by using a narrowband spectral filter to choose the fastest decay processes.



In conclusion we have demonstrated THz to optical conversion using an atomic vapour and presented real-time, near-field THz field images, with the potential for fast image acquisition in a 2D-plane.
Our method presents a significant benefit over THz field mapping where photoconductive antennae~\cite{mitr09} or electro-optic crystals~\cite{bitz08,seo07} are rastered, leading to image acquisition on the time scale of tens of hours.
Recent high-resolution work imaging THz surface waves still required around a 1 hour collection time despite claiming a significant speed-up~\cite{wang16}.
The caesium vapour does not absorb or otherwise distort the THz field, and so we are able to image narrowband THz waves traveling in opposite directions superposed as a standing wave.
This would be impossible using either detectors that absorb energy from the THz field such as pixel arrays~\cite{lee05}, or detectors that rely on short, high-intensity THz pulses. 
Traceability to SI units makes our technique a candidate for an atomic reference candle in the THz range~\cite{camp98}, yet fluorescence imaging gives a substantial advantage over EIT imaging because the resolution is not set by the geometry of the vapour cell~\cite{fan14}. 
Because our THz-to-optical conversion medium is in the gas phase we envisage immersing structures within the caesium vapour in order to measure diffraction and reflection around wavelength-scale obstructions.

\vspace{0.5cm}


\noindent{\bf Acknowledgments:} The authors would like to thank Mike Tarbutt, Andrew Beeby, Andrew Gallant and Claudio Balocco for the loan of equipment and acknowledge funding from Durham University, The Federal Brazilian Agency of Research (CNPq), and EPSRC (Grants EP/M014398/1 and EP/M013103/1). 
All data are available on request.

\vspace{1cm}
\noindent{\bf Methods}
\vspace{0.1cm}

\noindent{\bf Laser Excitation:} We use a three-step excitation process to excite caesium atoms to the Rydberg state. The probe laser (852\,nm) excites atoms to the 6P$_{3/2}$ $F$ = 5, state using 21~$\mu$W power (1/e$^2$ radius 30\,$\mu$m). The coupling laser (1470\,nm) takes the atoms from the 6P$_{3/2}~F$~=~5 state to the 7S$_{1/2}~F$~=~4 state and has 89~$\mu$W power (1/e$^2$ radius 90\,$\mu$m). Both the probe and coupling lasers are stabilised to the atomic resonances using polarisation spectroscopy~[31]. The Rydberg laser (799\,nm) is tuned to the 7S$_{1/2}$ to 21P$_{3/2}$ state transition but we do not resolve the hyperfine structure. The Rydberg laser is not frequency stabilised, and 520\,mW (1/e$^2$ radius 130\,$\mu$m) is required due to the much weaker excited-state atomic transition. All three laser beams are co-axial inside the vapour cell. The Rydberg laser propagates in the opposite sense to the probe and coupling beams.

\noindent{\bf Terahertz Beam:} The Terahertz beam is generated from a microwave signal using a frequency multiplier chain (\emph{WR1.5-AMC, Virginia Diodes Inc.}). We estimate that the total power in the beam is 10~$\mu$W, which is consistent with the electric field measured with the EIT. The spectral linewidth is below the MHz level and much narrower than the bare EIT line width (Figure~2a, green line) for which we fit a FWHM of 8 MHz in figure 1a (green line). The THz beam is linear polarised and matches the polarisation of the Rydberg laser beam. By matching the polarisations we avoid exciting 21P$_{3/2}$ m$_{\rm{j}} = \pm \frac{3}{2}$ states which are not coupled to the 21S$_{1/2}$ m$_{\rm{j}} = \pm \frac{1}{2}$ states by the THz field (See Supplementary Information). The standing wave is caused by a combination of reflection from the quartz cell containing the caesium vapour and the heating oven.

\noindent{\bf Atomic Vapour:} The caesium is contained in a quartz cell, with laser path length of 2~mm. The temperature of the vapour is stabilised using a stainless steel oven which encases the glass cell. For the images presented in this letter the vapour temperature was 68$^{\circ}$C. We found that temperatures around 70$^{\circ}$C provided sufficient vapour pressure for fast (0.5 second) fluorescence images without excessive absorption of the probe laser beam.

\noindent{\bf Camera Images:} The fluorescence images are recorded using a \emph {Canon} 5D MkIII consumer digital camera. The strong suppression of IR sensitivity in the camera suits our method as it removes unwanted contributions from scattered laser light and fluorescence from low lying atomic states. A 5X zoom, f/5.6 macro lens is used to image the light. The camera ISO was 25600. The RAW camera files are processed using \emph{Matlab} to ensure independence of the three colour channels. The video images were recorded using the same device, but due to file compression we are unable to linearise the pixel signal or separate the colour channels. The images were recorded with $\Delta/2\pi$~=~-243\,MHz, where negative~$\Delta$ denotes red detuning of the laser.

\noindent{\bf Fluorescence Spectrum:} We used an {\it Oriel~Cornerstone~130} monochromator and a {\it Hamamatsu~H10722.110} photon multiplier tube to measure the fluorescence spectrum. The spectral sensitivity of the combined system was calibrated against a known white light source and the wavelength selection of the monochromator was calibrated against caesium lines of known wavelength. The spectrum was recorded at 71$^{\circ}$C with $\Delta/2\pi$~=~-367\,MHz, taking around an hour to record.



\vspace{1cm}

\noindent [31] Carr,\,C., Weatherill,\,K.\ J.,and  Adams,\,C.\ S. 
{\rm Polarization spectroscopy of an excited state transition} 
{\it Opt. Lett.} {\bf 37} 118 (2012).

\end{document}